\documentclass[manuscript]{acmart}

\setcopyright{none}
\usepackage{booktabs}
\usepackage{graphicx}
\usepackage{multirow}
\usepackage{tabularx}
\usepackage{array}
\usepackage{xspace}

\newcolumntype{Y}{>{\raggedright\arraybackslash}X}

\title{From the Task Boundaries of Narrative Text to Structural Anchoring, Uncertainty Triggers, and Cross-Calibration}

\author{Bowen Deng}
\email{dengbw5@mail2.sysu.edu.cn}
\affiliation{%
  \institution{Sun Yat-sen University}
  \country{China}
}

\author{Jiaqi Zou}
\email{zoujq3@mail2.sysu.edu.cn}
\affiliation{%
  \institution{Sun Yat-sen University}
  \country{China}
}

\author{Kexin Zhang}
\email{zhangkx57@mail2.sysu.edu.cn}
\affiliation{%
  \institution{Sun Yat-sen University}
  \country{China}
}

\author{Daifeng Li}
\authornote{Corresponding author.}
\email{lidaifeng@mail.sysu.edu.cn}
\affiliation{%
  \institution{Sun Yat-sen University}
  \country{China}
}

\acmConference[CHI '27]{CHI Conference on Human Factors in Computing Systems}{2027}{Pittsburgh, PA, USA}



\begin{document}

\begin{abstract}
Causal graphs represent structural relationships among variables, yet users must still interpret direction, mechanism, and adjustment conditions in relation to the task at hand. Prior work often compares explanation formats as fixed conditions and pays less attention to how users distribute reasoning across graphs, direct explanations, and stories. We developed CoNS-Explorer, which uses reviewed instructional DAGs/SCMs to maintain a shared causal-fact ledger and generate fact-matched direct explanations and contextualized stories. A controlled survey experiment ($N=240$) compared the two texts as complete presentation packages. In the primary GLMM, the Story condition had a positive but uncertain overall association with accuracy (OR $=1.55$, 95\% CI $[0.34,7.10]$, $p=.572$); a population-averaged GEE showed a significant positive effect (OR $=1.89$, 95\% CI $[1.02,3.48]$, $p=.042$). Task-type interactions localized the clearest advantage to total-effect adjustment. Story also significantly increased situational presence. In a separate system-task and interview study ($N=24$), participants freely used graphs, direct explanations, and stories across three causal models. They established structural anchors with graphs and numerical results, consulted text when direction was unclear, mechanisms were unfamiliar, or multiple paths competed, and checked their judgments against other representations or external evidence. Integrating the two studies, we develop a process framework of structural anchoring, uncertainty triggering, explanation routing, and cross-calibration, together with four testable design propositions for adaptive causal explanation.
\end{abstract}

\maketitle

\section{Introduction}
\label{sec:introduction}

Causal graphs appear in a growing range of data-analysis and AI interfaces, where arrows summarize structural relationships among variables. A graph can show which variables are connected, but it does not automatically supply the explanation required by a particular task. Users must still determine the direction of a relationship, distinguish observation from intervention, identify common causes and common outcomes, and understand how controlling for a mediator changes the meaning of an effect estimate~\cite{hernan2020,pearl2009,sloman2015}. Causal graphs therefore provide a representative setting for human-facing explanation: the graph compresses relationships, language supplies task-relevant information, and users coordinate reasoning across the two.

A direct explanation is organized around variables, directions, and adjustment conditions, making local relationships easy to verify. A story places abstract variables within characters, goals, and a sequence of events, thereby turning them into an imaginable process. Narrative research shows that situated, temporally ordered accounts can support mental simulation and shape beliefs~\cite{green2000,dahlstrom2014,braddock2016}. Evidence from human-centered XAI also shows that readability, engagement, and judgment accuracy can change independently~\cite{bucinca2021,danry2023,li2026,tian2026,poursabzi2021}. Explanation design must consequently address both the information a representation provides for the current task and the point at which users consider that information worth its reading cost.

Recent narrative XAI systems have transformed SHAP outputs, counterfactual explanations, and graph-model explanations into natural-language stories~\cite{zytek2024,martens2025,cedro2026,lukassen2026}. Work on causal visualization likewise demonstrates the value of combining text with structured graphics~\cite{guo2023,yan2020,yen2021,choudhry2021}. Two connected questions remain. First, narrative and direct explanations often differ simultaneously in factual content, wording, and cue salience, which makes the task boundary of the narrative package difficult to identify under fact-matched conditions. Second, comparisons between fixed conditions do not capture how resources work together in realistic interaction. Users may begin with a graph, search a precise explanation, read a story around an unfamiliar mechanism, and then return to the graph or data to verify a judgment. The average effect of an explanation format and the process through which users actually invoke it therefore require a shared research design.

We developed CoNS-Explorer to study these questions. The system generates a variable-centered direct explanation and a contextualized story from the same causal-fact ledger. Both texts preserve the causal structure, direction, numerical values, intervention results, and model boundaries. We use a sequential explanatory mixed-methods design~\cite{fetters2013}. A between-subjects survey experiment with 240 participants estimates the overall effects of the two complete texts and explores differences across causal tasks. A separate system-task and interview study with 24 participants allows free access to the graph, direct explanation, and story, revealing what triggers a switch between information sources, what function each source serves, and how users verify a judgment.

We address three research questions:

\noindent\textbf{RQ1.} When core causal facts are held constant, how do Direct and Story texts affect task accuracy and post-task experience ratings?

\noindent\textbf{RQ2.} Does the effect of the text vary across confounding, outcome-selection, and total-effect-adjustment tasks?

\noindent\textbf{RQ3.} When graphs, Direct text, and Story text are simultaneously available, what local uncertainties trigger resource switching, and how do users verify judgments across representations?

The two studies provide complementary evidence. In the controlled experiment, Story had a positive but modest association with overall accuracy, with the clearest benefit in total-effect-adjustment tasks. Situational presence showed the most stable change among post-task ratings. The free-browsing study revealed a resource-selection process: participants established structure from the graph and intervention results, then consulted text when local uncertainty arose. Direct text helped confirm direction, boundaries, and local mechanisms, whereas Story helped participants understand processes in unfamiliar domains. Together, these results recast explanation design from choosing one globally preferred format to routing resources around a user's current reasoning gap.

This paper makes three contributions:

\begin{itemize}
  \item We implement a fact-matched dual-text design in CoNS-Explorer. Direct and contextualized Story texts share a DAG/SCM, numerical values, intervention results, and model boundaries, providing an auditable basis for comparing generated explanations.
  \item We provide boundary evidence on the average effect and task heterogeneity of Story text through a controlled study ($N=240$), then use an independent free-browsing study ($N=24$) to explain how graphs, Direct text, and Story text divide explanatory work. This connects evidence from fixed comparisons with observed use processes.
  \item We synthesize the two studies into a process framework spanning structural anchoring, uncertainty triggering, explanation routing, and cross-calibration. Four observable propositions guide interface implementation and evaluation.
\end{itemize}

\section{Related Work}
\label{sec:related}

\subsection{Human-Centered XAI Requires Task-Based Validation}

Human-centered XAI has long argued that explanations should be organized around users' questions, goals, and actions and evaluated at the task level. Transparency into model internals is only one component of explanation quality. Lim et al. compared why and why-not explanations~\cite{lim2009}; Liao et al. organized XAI requirements around questions users ask~\cite{liao2020}; Miller characterized explanations as contrastive, selective, and social~\cite{miller2019}; and Wang et al. translated related principles into a design framework~\cite{wang2019}. Kulesza et al. connected explanation effectiveness with the soundness and revisability of users' mental models~\cite{kulesza2012,kulesza2015}, while Poursabzi-Sangdeh et al. showed that greater transparency does not necessarily improve error detection~\cite{poursabzi2021}. In human--AI decision making, explanations may even increase reliance on both correct and incorrect advice~\cite{bansal2021,kim2025,schemmer2023,raees2026,bo2025,bucinca2021}. These findings make task-valid understanding distinct from ratings such as liking, clarity, or trust. We therefore treat objectively scored causal-discrimination performance as the primary outcome of Study 1 and report subjective experience separately.

\subsection{From Structured Explanations to Stories}

Narratives organize abstract relationships through characters, goals, and connected events. Narrative visualization research has likewise treated storytelling as a fundamental means of organizing and communicating insights from data~\cite{segel2010,boy2015,lan2021,kosara2013}. Green and Brock describe transportation as immersion involving attention, imagery, and emotional engagement~\cite{green2000}. Dahlstrom argues that narratives can lower barriers for nonexpert science audiences while also increasing persuasion through the realism of individual cases~\cite{dahlstrom2014}; Braddock and Dillard's meta-analysis further shows effects on beliefs and behavioral intentions~\cite{braddock2016}. In XAI, Explingo, XAIstories, and GraphXAIN explore narrative presentations of feature attribution, counterfactual, and graph neural-network explanations~\cite{zytek2024,martens2025,cedro2026}. Lukassen et al. demonstrate that explanation quality depends jointly on the model, the XAI method, and the generation strategy~\cite{lukassen2026}. This literature establishes stories as a viable explanation medium while exposing a key confound: characters, event order, contextual detail, and cue salience often change together. Our fact-matching framework supplies a tighter boundary for comparing complete texts while preserving their characteristic linguistic cues.

\subsection{Causal Visualization and Interactive Causal Reasoning}

Visualization systems can help users externalize causal structures, inspect counterfactuals, and explore interventions~\cite{guo2023,yan2020,borland2024,yen2021}. CAUSEWORKS and related research further combine textual narratives with causal visualizations to support relation recovery and intervention exploration~\cite{choudhry2021}. Causal-reasoning difficulty depends both on visible structure and on a variable's role in the current question: a common cause can change an observed association, selection on a common outcome can introduce a noncausal association, and adjustment for a mediator changes the estimated effect~\cite{hernan2020,pearl2009}. A presentation may thus help with one structural operation while contributing little to another.

Prior work establishes that explanations need task-based validation, stories can organize abstract relationships, and causal graphs can support structural exploration. We focus on how multiple explanation resources work together. The controlled experiment estimates where the two complete texts help within a fixed task set; the free-browsing study describes how users select, switch, and verify resources. Their integration addresses what help each resource provides, when users invoke it, and how it works with other evidence.

\section{CoNS-Explorer: Fact-Matched Causal Explanation}
\label{sec:system}

\subsection{Comparing Expressions Under Fact Consistency}

CoNS-Explorer compares and combines explanation formats for a given, reviewed local DAG/SCM. Its scope is the communication of causal relationships already represented in the model. We define an instructional material package as
\begin{equation}
  M=\{G,S_0,a,S_1,\Delta,B,Q\},
\end{equation}
where $G$ is the DAG/SCM, $S_0$ is the baseline state, $a$ is a fixed intervention, $S_1$ and $\Delta$ are the post-intervention state and change, $B$ is the model boundary, and $Q$ is the common question set. Direct text is generated as $D=f_D(M)$, while Story text is generated as $T=f_T(M,C)$, where $C$ contains only fictional characters, settings, goals, and temporal progression. Core edges, directions, baseline values, intervention values, updated outcomes, and boundaries in both texts originate from the same $M$.

Core-fact matching still permits each text to retain its characteristic linguistic cues. Direct text more readily uses terms such as ``control,'' ``common cause,'' and ``increase/decrease.'' Story may make particular relationships salient through temporal order and context. These differences jointly constitute each complete text. Study 1 therefore estimates the difference between the two presentation packages as a whole. Future factorial studies can isolate the contribution of individual components.

\subsection{Two Explanation Formats}

Direct text uses variables and statistical relationships as grammatical subjects and explicitly states directions, common causes, common outcomes, mediator positions, pre/post-intervention values, and model boundaries. Its goal is compact, searchable verification. A contextualized story must contain at least five narrative elements: a character or analyst clearly identified as fictional, a practical goal, a baseline situation, a valid trigger, and a sequence of changes following directed relationships in the graph. It closes by restating the instructional model boundary. A text that merely adds a person's name and continues to recite edges remains a direct restatement rather than a story.

\begin{table}[t]
\centering
\caption{Length matching between Direct and Story texts in Study 1. Character-count differences remained below 10\% for every pair.}
\label{tab:length}
\small
\begin{tabular}{lrrr}
\toprule
Material & Direct & Story & Difference \\
\midrule
Education, pre-intervention & 403 & 423 & 5.0\% \\
Education, post-intervention & 304 & 328 & 7.9\% \\
Macroeconomics, pre-intervention & 441 & 465 & 5.4\% \\
Macroeconomics, post-intervention & 309 & 339 & 9.7\% \\
\bottomrule
\end{tabular}
\end{table}

\subsection{Generation, Execution, and Verification}

The system operates in six steps. It first checks that the DAG is acyclic and validates node units, admissible values, and model version. It then executes the fixed intervention and updates only descendants specified by the model. Third, it writes relations, directions, pre/post values, unchanged variables, and boundaries to a shared fact ledger. Fourth, it generates either Direct or Story text from this ledger. Fifth, deterministic rules check entities, numerical values, directions, fact coverage, and unsupported causal claims. Finally, researchers assess whether the story reads naturally and whether its fictional context has been misrepresented as a real-world guarantee. Materials enter the study only after all critical checks pass.

\begin{figure}[t]
  \centering
  \includegraphics[width=\linewidth]{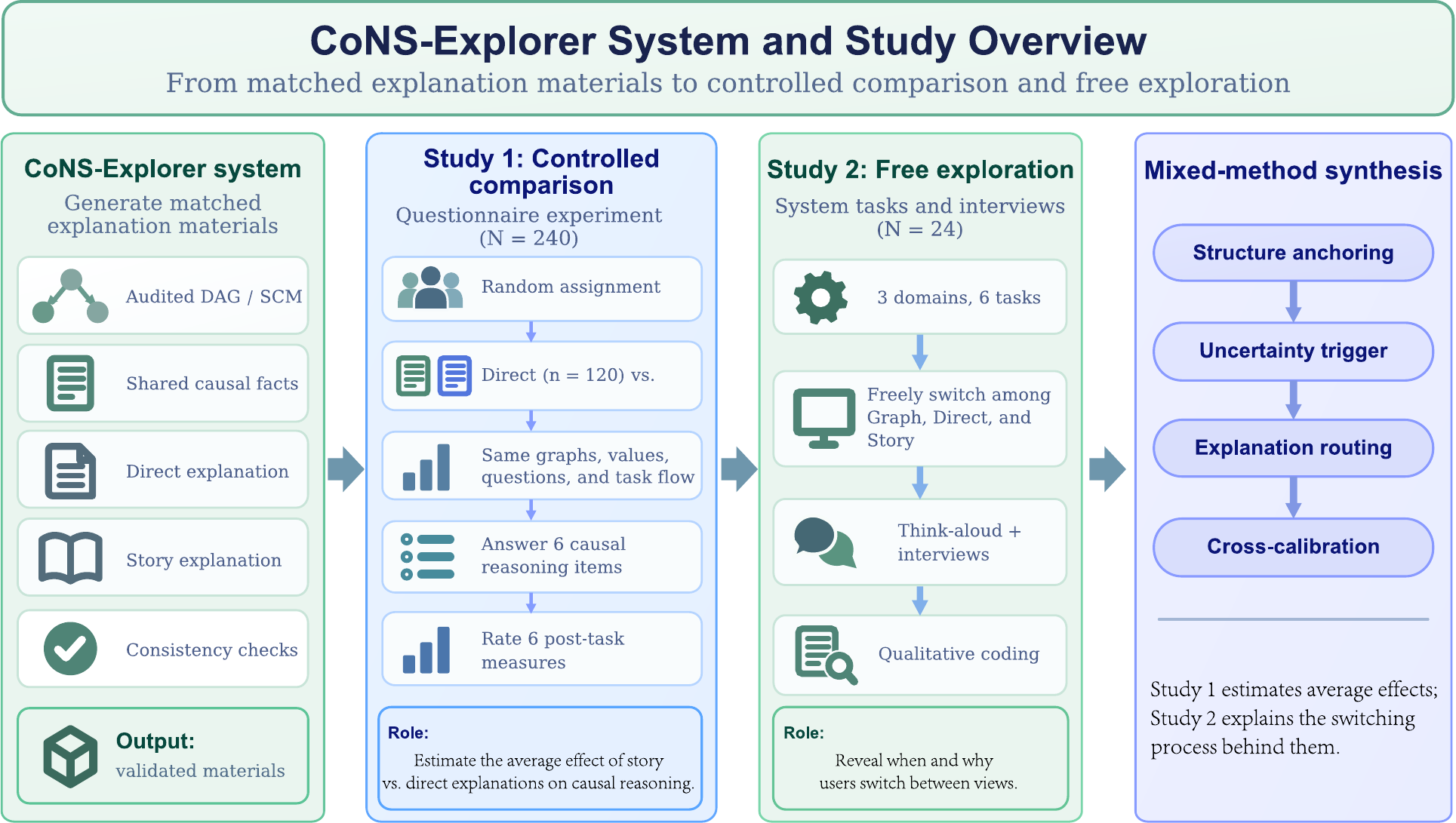}
  \caption{CoNS-Explorer and the two-stage mixed-methods design. The system generates fact-matched Direct and Story texts; Study 1 estimates the average effect of Story text, Study 2 explains free switching among resources, and their integration yields the explanation-routing process framework.}
  \Description{A workflow diagram links reviewed causal models to a shared fact ledger and two text formats. It then branches to a controlled survey experiment and a free-browsing interview study before converging on an explanation-routing framework.}
  \label{fig:overview}
\end{figure}

CoNS-Explorer communicates a supplied causal model. Inputs containing only a DAG support structural descriptions, whereas new post-intervention values require an executable SCM. The appendix documents the fact-ledger structure, generation and verification controls, task-to-answer mapping, statistical robustness checks, and qualitative-analysis workflow. Complete text pairs, line-level coding, the codebook, and the participant-by-code matrix are retained as separate anonymized companion artifacts. Together, these materials support tracing claims from model facts to study evidence. Figure~\ref{fig:overview} summarizes the workflow from fact-matched material generation through controlled comparison, free browsing, and mixed-methods integration.

\subsection{Study Materials and Causal Scenarios}

The education scenario concerns private tutoring and post-test performance. Family socioeconomic status, parental education, school quality, and pre-test scores provide background factors. Tutoring also affects self-directed learning, sleep, and test anxiety, which influence post-test scores through mediated and direct pathways. This model supports three forms of causal discrimination: overall versus stratified association, selection on an outcome, and mediator adjustment. The fixed intervention changes tutoring from 5.23 to 10.00 hours per week. The appendix summarizes the task-to-structure mapping and answer key in Table~\ref{tab:app-items} and documents a participant-facing education-system state in Figure~\ref{fig:app-screenshots}.

The macroeconomic scenario concerns monetary-policy transmission. The output gap and inflation expectations jointly affect the policy rate and continue to shape financing and inflation conditions. The policy rate operates through bank lending rates, credit, and investment to affect core inflation and unemployment. The fixed intervention changes the policy rate from 2.73\% to 5.40\%. This model supports reasoning about price-puzzle-style confounding, selection bias induced by restricting observations to high-policy-rate years, and adjustment-set choice for total effects. Its three task targets and prespecified answers are summarized alongside the education tasks in Table~\ref{tab:app-items}.

\section{Study 1: Controlled Survey Experiment}
\label{sec:study1}

Study 1 addresses RQ1 and RQ2. Overall correctness across six objective questions is the primary outcome for estimating the average effect of Story text over the predefined task set. Interactions and within-type contrasts characterize task heterogeneity. Six post-task ratings provide complementary evidence about experiential differences between the two texts.

\subsection{Design and Participants}

We used a two-group between-subjects design and randomly assigned participants 1:1 to a Direct (D) or Story (S) condition. The formal online study was conducted on the Jianshu platform on September 2, 2026. We received 240 complete responses, with 120 participants in each condition. Each participant received CNY 3, and no participant was excluded from analysis. Participants completed both the education and macroeconomic scenarios under the same text condition. Eligibility required participants to be at least 18 years old, fluent readers of Chinese, and able to understand correlations, stratified comparisons, and the basic meaning of ``controlling for a variable.'' Formal training in causal inference was not required.

The study protocol was reviewed and approved by the local academic committee as low-risk survey and interview research. All analyses and reporting used anonymized data. Survey participants read an informed-consent statement on the first page and indicated their consent. Interview participants in Study 2 again provided oral confirmation of their willingness to participate and authorization for audio or screen recording before beginning the tasks.

\subsection{Procedure and Causal-Discrimination Tasks}

Participants first examined the causal graph, baseline state, and assigned text for one scenario, then answered three multiple-choice questions without feedback. After the fixed intervention, they saw updated numerical values and a new text in the same condition before completing the remaining judgments. The procedure was repeated for the second scenario. Both groups received the same graphs, baseline states, interventions, updated values, and questions. Only the text format differed. The public study interface did not display the answer key or researcher explanations. Appendix Table~\ref{tab:app-items} summarizes the task-to-structure mapping and correct answers, while Appendix~\ref{app:materials} reports the fact- and length-matching checks for the paired texts.

\subsection{Data Preparation and Quality Control}

Before hypothesis testing, we audited all 240 complete responses, including 120 in each condition. The audit covered completeness, uniqueness of respondent and user IDs, repeated IP addresses, platform-recorded accuracy and duration, and questionnaire quality-control fields. All six objective responses and all six post-task experience ratings were complete, and respondent and user IDs were unique. A small number of repeated IP addresses were retained as audit flags only. The Direct questionnaire contained one additional attention-check item; because this item was asymmetric across conditions, it and completion duration were used only for quality auditing and did not enter the analysis. All 240 participants were retained in the primary analysis.

The six questions were scored as correct (1) or incorrect (0) using answers specified in advance from the reviewed DAGs/SCMs. We transformed participant-level records into an item-level long table with 1,440 participant--item observations. Each item was classified by its principal structural requirement as confounding, outcome selection, or total-effect adjustment, with two questions in each category. The item-level mixed-effects model provided the confirmatory test, and each participant's total score from 0 to 6 supported robustness analyses. Appendix~\ref{app:study1-processing} reports the data-quality audit and measurement decisions; Appendix~\ref{app:study1-results} reports the confirmatory and robustness estimates; and Table~\ref{tab:app-items} provides the task mapping and prespecified answer key.

\subsection{Outcomes and Statistical Analysis}

The six objective items formed a task set covering three causal-reasoning operations~\cite{devellis2022}. Correct answers were determined by the corresponding instructional DAG/SCM and prespecified structural criteria. Task validity was supported by the mapping between each question and its causal structure, fact alignment across text conditions, and the material-verification procedure.

Overall correctness across the six questions was the sole confirmatory primary outcome. We fitted a binomial logistic mixed-effects model:
\begin{equation}
\operatorname{logit}\{P(Y_{ij}=1)\}=\beta_0+\beta_1\operatorname{Story}_i+\gamma_j+u_i,
\label{eq:glmm}
\end{equation}
where $\gamma_j$ controls fixed differences among the six reviewed items and $u_i$ is a participant random intercept. We report the conditional odds ratio (OR) for Story relative to Direct, its 95\% confidence interval, and population-marginal predicted accuracy obtained by integrating over the random-effect distribution. All GLMMs were checked for optimization convergence and singular fit. Inference is bounded to the six tasks used in this study.

We assessed sensitivity to model specification with three complementary analyses. At the participant level, we compared six-item total scores and reported a bootstrap confidence interval and Hedges' $g$. A participant-cluster bootstrap re-estimated overall marginal accuracy and the Story--Direct difference. A GEE with participant-clustered robust sandwich standard errors estimated the population-averaged effect. The GLMM estimates a conditional effect after accounting for participant-level heterogeneity, whereas GEE estimates a population-averaged effect; we report both estimands. After freezing the confirmatory model, we tested the Story Condition $\times$ Task Type interaction and applied Holm correction to the three within-type comparisons.

After both scenarios, participants rated vividness, mental simulation, situational presence, narrative transportation, trust within the instructional model, and cognitive effort on seven-point scales~\cite{hart1988}. Each construct was measured by a single item and represents a post-task experiential rating. The first four outcomes were analyzed with cumulative-logit models and Holm correction. HC3 robust linear models for all six outcomes provided sensitivity checks.

\subsection{Results}

\subsubsection{Boundary of the Overall Effect}

Participants performed well overall. Raw mean accuracy was 87.9\% in Direct and 91.5\% in Story; 76.7\% of participants answered all six questions correctly, and pooled accuracy was at least 90\% for four items. In the confirmatory GLMM, the point estimate favored Story but remained uncertain (OR $=1.55$, 95\% CI $[0.34,7.10]$, $p=.572$). High accuracy constrained the room for differences on the binary outcome and contributed to the wide interval~\cite{devellis2022}. After integration over the random-effect distribution, predicted marginal accuracy was 86.1\% for Direct and 87.2\% for Story, a difference of $+1.16$ percentage points (participant-cluster bootstrap 95\% CI $[-0.48,3.25]$).

Complementary analyses produced estimates in the same direction. The overall GEE showed a significant positive effect of Story on accuracy (OR $=1.89$, 95\% CI $[1.02,3.48]$, $p=.042$). Participant total scores were also higher in Story (5.49 vs. 5.28; Hedges' $g=0.161$). Together, these estimates indicate a positive effect on accuracy of limited magnitude.

\begin{figure}[t]
  \centering
  \includegraphics[width=\linewidth]{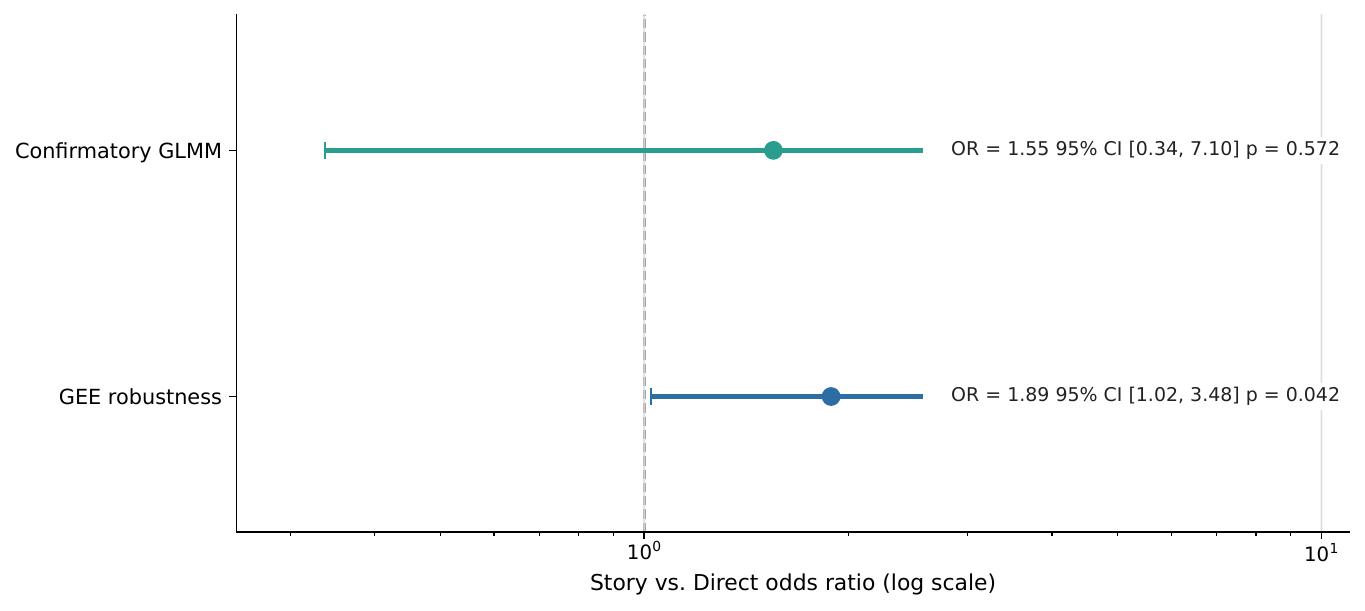}
  \caption{Overall effect of Story relative to Direct. Point estimates from both the GLMM and GEE favor Story; only the population-averaged GEE reaches statistical significance.}
  \Description{A two-row forest plot compares the odds ratios from a mixed-effects model and a generalized estimating equation. Both estimates exceed one, while the mixed-effects confidence interval is wider and crosses one.}
  \label{fig:overall}
\end{figure}

\subsubsection{Story Effects Varied Across Task Types}

The Story Condition $\times$ Task Type interaction appeared in both analytic frameworks (GLMM likelihood-ratio $\chi^2(2)=8.06$, $p=.018$; GEE robust Wald $\chi^2(2)=7.02$, $p=.030$). Cross-model localization showed that the Story effect was stronger for total-effect adjustment than for confounding. The ratio of effect odds ratios was 7.18 in the GLMM (Holm $p=.037$) and 2.21 in the GEE (Holm $p=.049$). The contrast between total-effect adjustment and outcome selection survived correction in the GLMM. The consistent direction across the two frameworks identifies total-effect adjustment as the task type with the strongest Story effect.

Within-type results supported this interaction. In the GEE, Story increased accuracy on total-effect-adjustment questions by 9.08 percentage points (95\% CI $[3.06,15.10]$, Holm $p=.009$), corresponding to OR $=3.17$ (Holm $p=.021$). The GLMM contrast pointed in the same direction but had a wider interval (OR $=6.82$, Holm $p=.166$). Total-effect adjustment therefore provides the clearest evidence of a Story benefit and is the highest-priority target for preregistered replication with a larger set of parallel items.

\begin{figure}[t]
  \centering
  \includegraphics[width=\linewidth]{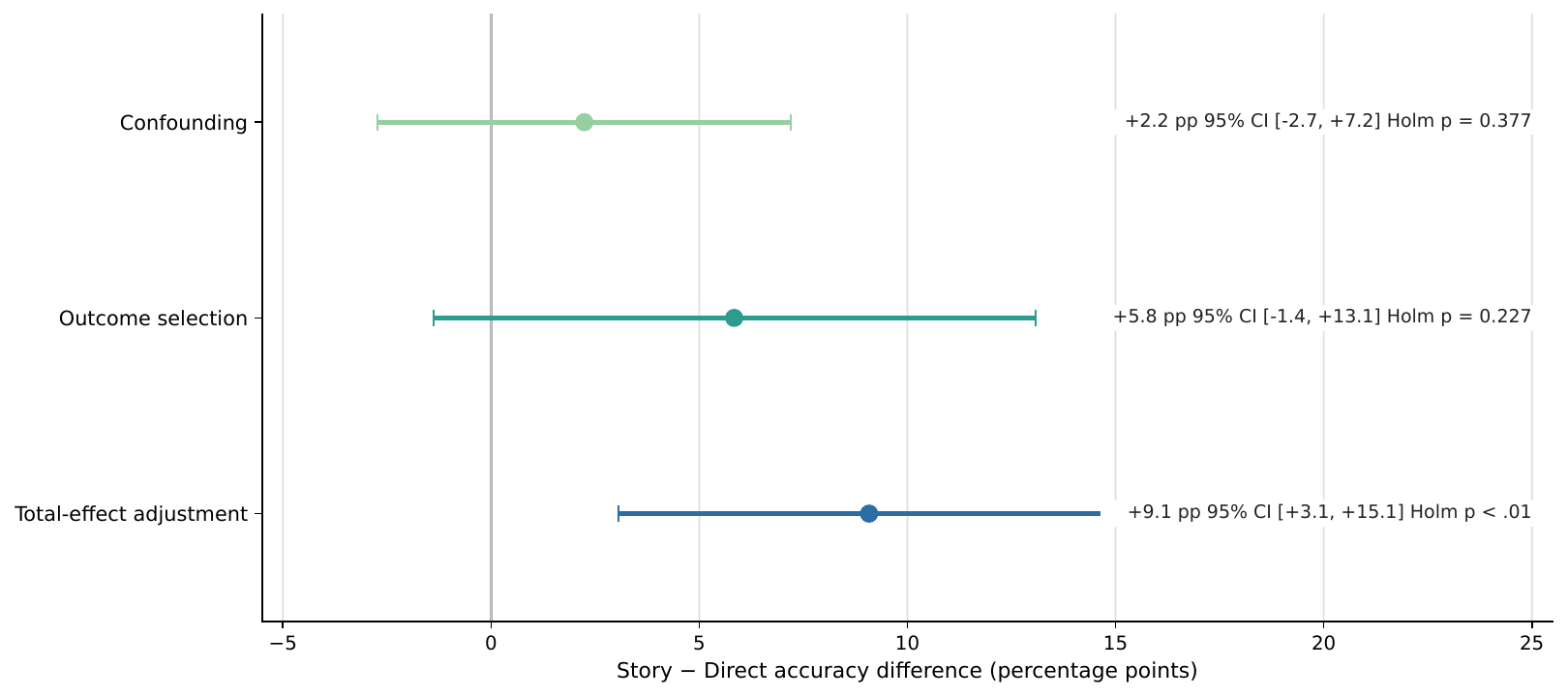}
  \caption{Exploratory GEE estimates of Story--Direct accuracy differences across the three causal task types.}
  \Description{A three-row coefficient plot shows Story-minus-Direct accuracy differences for confounding, outcome selection, and total-effect adjustment. The largest positive difference is for total-effect adjustment.}
  \label{fig:heterogeneity}
\end{figure}

\subsubsection{Story Significantly Increased Situational Presence}

Among the six post-task ratings, Story significantly increased situational presence. Mean ratings were 5.22 in Direct and 5.77 in Story; the cumulative-logit model estimated OR $=2.95$, 95\% CI $[1.81,4.79]$, Holm $p<.001$. The HC3 sensitivity analysis produced the same conclusion (difference $=+0.55$, 95\% CI $[0.32,0.78]$). Vividness, mental simulation, and narrative transportation were similar across conditions. Trust within the model was slightly lower in Story (5.57 vs. 5.73; OR $=0.66$, $p=.092$), and cognitive effort was nearly unchanged (4.88 vs. 4.94; OR $=0.93$, $p=.747$).

The experiential effect of Story was therefore concentrated in a sense of being situated in the scenario rather than a broad increase in subjective ratings. Study 2 examines when participants actually consult Story and when they rely on Direct text for precise verification.

\begin{figure}[t]
  \centering
  \includegraphics[width=\linewidth]{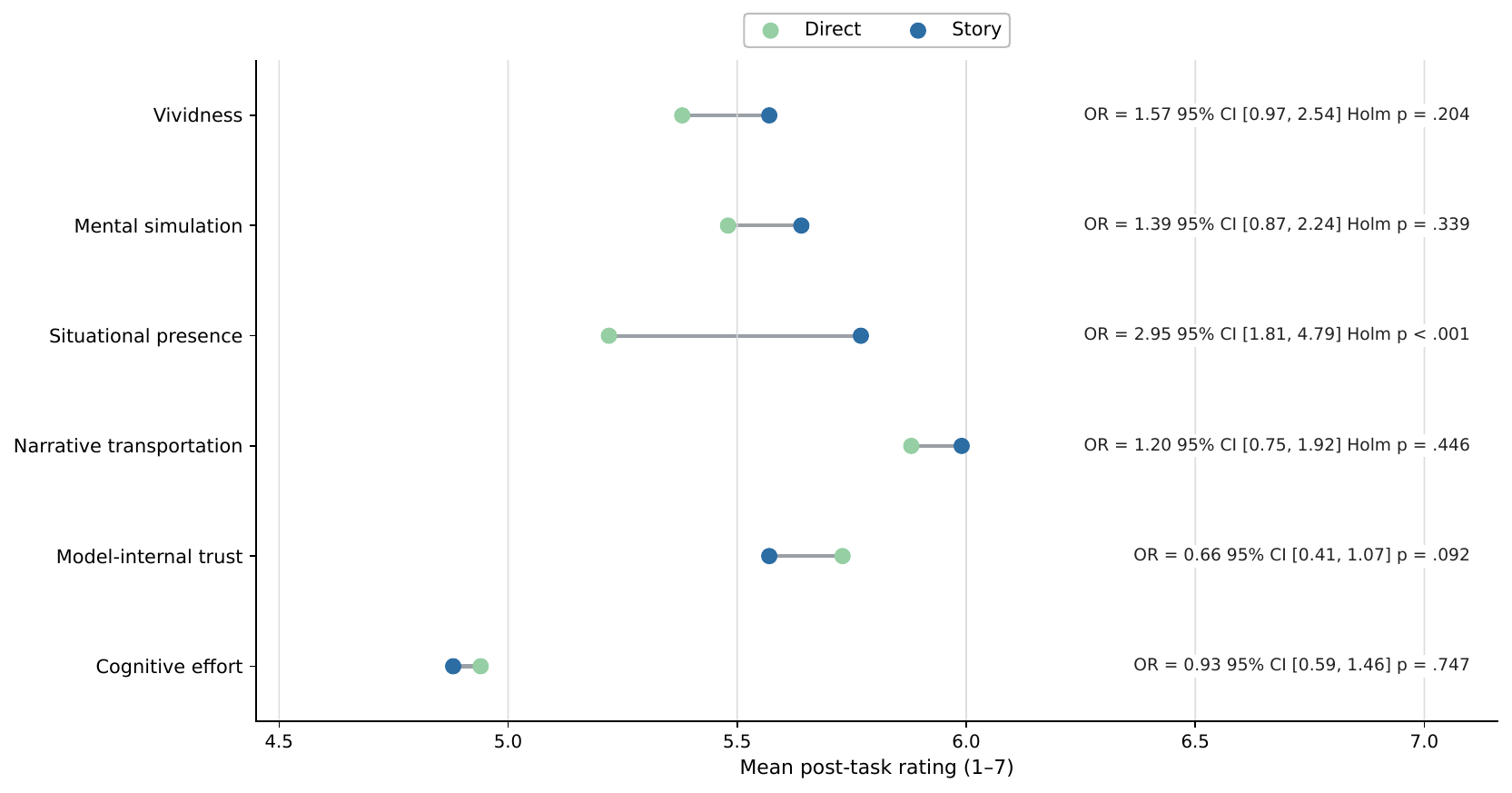}
  \caption{Effects of Story text on six post-task experience ratings. Situational presence shows the only statistically reliable increase.}
  \Description{A six-row forest plot presents odds ratios for vividness, mental simulation, situational presence, transportation, trust, and cognitive effort. The situational-presence interval lies clearly above one; the remaining intervals overlap one.}
  \label{fig:experience}
\end{figure}

\section{Study 2: System Tasks and Interviews}
\label{sec:study2}

Study 2 addresses RQ3 and provides a process account of the effect patterns observed in Study 1. Study 1 fixed explanation format between participants to estimate the task effect of each complete text. Study 2 made the graph, Direct text, and Story text simultaneously available and allowed participants to choose among them. The two studies used independent samples and are integrated through a joint display in Section~\ref{sec:discussion}.

\subsection{Participants and Research Setting}

After locking the Study 1 dataset, we recruited through WeChat 24 participants who had not taken part in the survey. Data collection ran from September 3 to 7, 2026, ensuring independent samples. Participants were over 20 years old and had at least an undergraduate education. They reported no or only introductory knowledge of causal inference, statistics, or data analysis. Each received CNY 20.

Participants used CoNS-Explorer to complete six interactive tasks. At any time, they could switch freely among three information sources: an interactive causal graph with node values, professional/direct text, and a contextualized story. The analysis sought to identify reasoning difficulties, reasons for switching resources, and verification practices after a switch.

Sessions were conducted either through Tencent Meeting with screen recording or in person with audio recording. All participants used a PC, and a complete system-task and interview session lasted approximately 40 minutes on average. Before beginning, participants read the study information and orally confirmed their willingness to participate and their authorization for recording. We anonymized all interview material before analysis and quotation; the paper identifies participants only as P01--P24.

\subsection{Three Domains and Six Interactive Tasks}

To broaden the contexts in which participants selected explanations, we added rice nitrogen management and the supply-chain bullwhip effect to the education scenario. Each causal model supported two tasks. In the rice domain, participants reasoned about how yield changes as nitrogen application continues to increase and how fixing the lodging rate changes the meaning of a total effect. Education tasks concerned downstream and upstream changes after directly setting self-directed learning, and the role of confounding in an observed pattern where students receiving more tutoring have lower average scores. Supply-chain tasks concerned the protective effect and potential cost of safety stock, and how fixing mean lead time changes the target effect. Across the six tasks, the designs covered mediator chains, competing positive and negative paths, the distinction between observation and intervention, and mediator adjustment.

Tasks appeared in a fixed order. Participants manipulated intervenable nodes, observed changes in color and value, and submitted a judgment. Within each task, they decided which information source to consult first and whether to open one or both texts. Researchers first used retrospective prompts tied to participants' observed actions: what they examined first, where hesitation emerged, why they changed information sources, and how they checked the resulting judgment. After this behavioral reconstruction, interviewers separately asked whether Story supplied mechanism information and when and at what granularity an ideal LLM explanation should appear. We marked evidence from observed behavior, retrospective reconstruction, and design probes separately during analysis.

\subsection{Qualitative Analysis}

AI-assisted transcription produced an initial transcript of each audio or screen recording. A researcher then checked the transcript against the original recording, and only verified, anonymized text entered the analysis. The codebook, coded excerpts, and participant-by-code matrix were maintained in Excel.

An LLM generated candidate initial codes. The first researcher returned to the anonymized source text line by line, removing codes that departed from participants' intended meaning, revising labels, merging duplicates, and adding omitted codes. The researcher consolidated 20 initial codes into six axial categories, compared cases through a participant-by-code matrix using the Framework Method~\cite{gale2013}, and manually developed 10 themes. Selective coding then organized the themes around structural anchoring, uncertainty triggering, explanation routing, and cross-calibration.

A second researcher independently rechecked the alignment between the anonymized source text and final codes, code boundaries, thematic assignments, quotations, and participant counts. When disagreements arose, both researchers returned to the original passage and discussed it until reaching a shared judgment. We retained cases that diverged from the main pattern, including participants who did not use Story, found Story insufficient for a judgment, or described it as distracting. The LLM was limited to proposing candidate initial codes; axial analysis, theme development, and all analytic decisions were performed by the researchers. Cross-checking and repeated comparison with the source text preserved traceability throughout the coding chain. Table~\ref{tab:coding-chain} illustrates representative coding paths. The complete line-level coding, codebook, and participant matrix are provided in the accompanying workbook, with observed behavior distinguished from preferences elicited by design probes.

\begin{table}[t]
\centering
\caption{Representative coding chains from the interview analysis. Source expressions and participant IDs are traceable in the accompanying coding workbook.}
\label{tab:coding-chain}
\scriptsize
\begin{tabularx}{\linewidth}{>{\raggedright\arraybackslash}X >{\raggedright\arraybackslash}p{0.21\linewidth} >{\raggedright\arraybackslash}p{0.18\linewidth} >{\raggedright\arraybackslash}p{0.16\linewidth}}
\toprule
Source expression or behavior & Initial code & Axial category & Process position \\
\midrule
``The image helps my causal inference more; it is more intuitive than text'' (P16) & Graph/action first; trace arrows & Structured exploration & Structural anchoring \\
``I don't know whether the arrow is positive or negative; when unsure, I check the explanation'' (P02) & Direction check & Mechanism completion & Precision routing \\
``I first get a preliminary idea, then corroborate it with the professional explanation'' (P11) & Story to Direct; cross-view check & Uncertainty-driven switching & Trigger and check \\
When Direct text was long, read only key information near the end (P07) & Long-text burden; selective reading & Cognitive economy & Cost modulation \\
When the system conflicted with experience, searched for third-party evidence (P10) & Challenge; external evidence & Epistemic calibration & Cross-calibration \\
``Divide it into a short explanation and a detailed explanation'' (P09) & Progressive disclosure & Adaptive AI explanation & Design requirement \\
\bottomrule
\end{tabularx}
\end{table}

\subsection{Interview Findings}

We present findings in the order participants generally followed when reasoning. They began by anchoring structure in the graph and intervention results. When this structural information did not support a judgment, they consulted Direct or Story text for different forms of information. Local uncertainty triggered a switch, while familiarity and reading cost altered the switching threshold. Competing paths or conflict with prior experience prompted further verification across representations. This sequence provides the empirical basis for the process framework in Section~\ref{sec:routing-framework}.

\subsubsection{Reasoning Began with Structural Anchors in the Graph and Intervention Results}

Participants commonly established a structural skeleton by ``changing one node, seeing which nodes changed, and tracing the arrows.'' All 24 participants used at least one of the graph, interaction, color, numerical value, or arrow as a structural anchor. Their starting points still varied: some first browsed Story or Direct text and then returned to the graph to organize relationships. P16 summarized the graph's core value as an intuitive representation of parent nodes, child nodes, and connected chains:

\begin{quote}
``This image actually helps my causal inference more. Compared with text, it is more intuitive.'' (P16)
\end{quote}

These anchors also exposed the functional boundary of the graph. Arrows primarily showed that relationships existed. Additional information was still needed to determine positive or negative direction, explain why a relation held, or identify the dominant relationship when multiple paths changed simultaneously. Twenty of 24 participants used Direct text to verify direction, conditions, or local logic; 23 of 24 either used Story or explicitly recognized its value for contextual or mechanistic understanding.

\subsubsection{The Two Texts Supplied Different Missing Information}

The graph generally showed connectivity, while Direct text stated directions, adjustment conditions, and local mechanisms. Participants searched for variable names, terms such as ``increase/decrease'' and ``positive/negative,'' and phrases indicating control conditions, aligning the text with the graph and wording of the question. P02 explained that Direct text became useful when an arrow's direction was unclear:

\begin{quote}
``I don't know whether this arrow has a positive or negative effect. When I'm not sure, I go to the explanation.'' (P02)
\end{quote}

Some participants read only the concluding part of Direct text. Its value therefore combined professional precision with searchability.

Story helped explain \emph{why} unfamiliar variables were connected. P24 initially did not understand the relationship between leaf-area index and lodging rate. Story organized the connection into a continuous process in which a larger leaf area made lodging more likely, supporting a judgment. Story also served as an entry point to unfamiliar terms or domains. This function varied across individuals: P04 relied on the graph and professional explanation throughout the tasks and did not consult Story.

\begin{quote}
``At first I didn't really understand leaf-area index or lodging rate. After reading the vivid story, I understood that when the area becomes too large, lodging becomes more likely.'' (P24)
\end{quote}

The ordering of the two texts depended on the current information gap. In the supply-chain task, P19 reported that unfamiliarity made Story alone insufficient and that the summary in Direct text was still needed to confirm how variables changed. Story supported mechanistic intuition, Direct text stated verifiable relationships, and the graph and numerical results provided a shared basis for checking both.

\subsubsection{Local Uncertainty Triggered Switching, Modulated by Familiarity and Reading Cost}

The division of labor among resources appeared as dynamic switching. Twenty of 24 participants described switches triggered by local uncertainty. They moved from Direct to Story when professional text was too long, terminology was difficult, or target information was not immediately visible. They returned from Story to Direct when Story omitted an explicit relation, appeared insufficiently rigorous, or required confirmation. P11 summarized this sequence as forming a preliminary idea from Story and then corroborating it with Direct text:

\begin{quote}
``After reading the story, I have a preliminary idea, and then I need to corroborate it with the professional explanation.'' (P11)
\end{quote}

Whether a local gap prompted a switch also depended on domain familiarity and reading cost. Fifteen of 24 participants explicitly linked familiarity to information choice. In the familiar education scenario, they relied more on the graph, values, and existing knowledge. Unfamiliar agriculture and supply-chain settings generated greater demand for text. Twenty-two of 24 participants described burdens associated with length, repetition, terminology, or uninterrupted prose; they reduced this cost through keyword search, reading only the final paragraph, or looking for the conclusion first. Seven participants explicitly reported that characters, plot, or a story-like style introduced irrelevant detail or distraction.

\begin{quote}
``The story context is too strong. It brings in other information about the situation that isn't necessary.'' (P24)
\end{quote}

Switching thus reflected a tradeoff among local informational value, credibility, and reading cost. Participants looked for the smallest amount of information sufficient to resolve the current uncertainty and expanded the explanation when needed.

\subsubsection{Competing Paths Prompted Cross-Checking and Requests for Controls}

When an outcome was shaped by several positive and negative paths, structural anchors alone often failed to establish which relation dominated. Seventeen of 24 participants mentioned competing paths, multiple parents, nonlinear trends, or the need to control variables. P01 wanted the system to state how A would affect B with all other variables held constant, while P16 requested a single-variable line chart to show the controlled trend. These requests concern local path isolation and trend comparison; additional narrative detail would not directly supply the requested operation.

\begin{quote}
``Tell me how a change in factor A would affect factor B when the other variables remain unchanged.'' (P01)
\end{quote}

Participants actively checked system explanations. Fourteen of 24 calibrated judgments among the graph, Direct text, Story, real-world experience, or external evidence. When model output conflicted with experience, they searched for additional information, inspected the data source, sought third-party evidence, or questioned both the system and their own intuition. Effective explanations therefore need to state a conclusion while preserving paths back to structure, values, sources, and alternative representations.

In direct probes about ideal LLM explanations, 21 of 24 participants expressed a preference for layered, on-demand, or dynamic presentation. They wanted a graph or short conclusion first, with Direct or Story details expanded when they encountered difficulty. Some requested continuous-change displays, trend lines, or the currently active intervention chain. These responses came from design probes and represent stated preferences rather than observed switching behavior.

\subsection{An Explanation-Routing Process Framework}
\label{sec:routing-framework}

\begin{figure}[t]
  \centering
  \includegraphics[width=\linewidth]{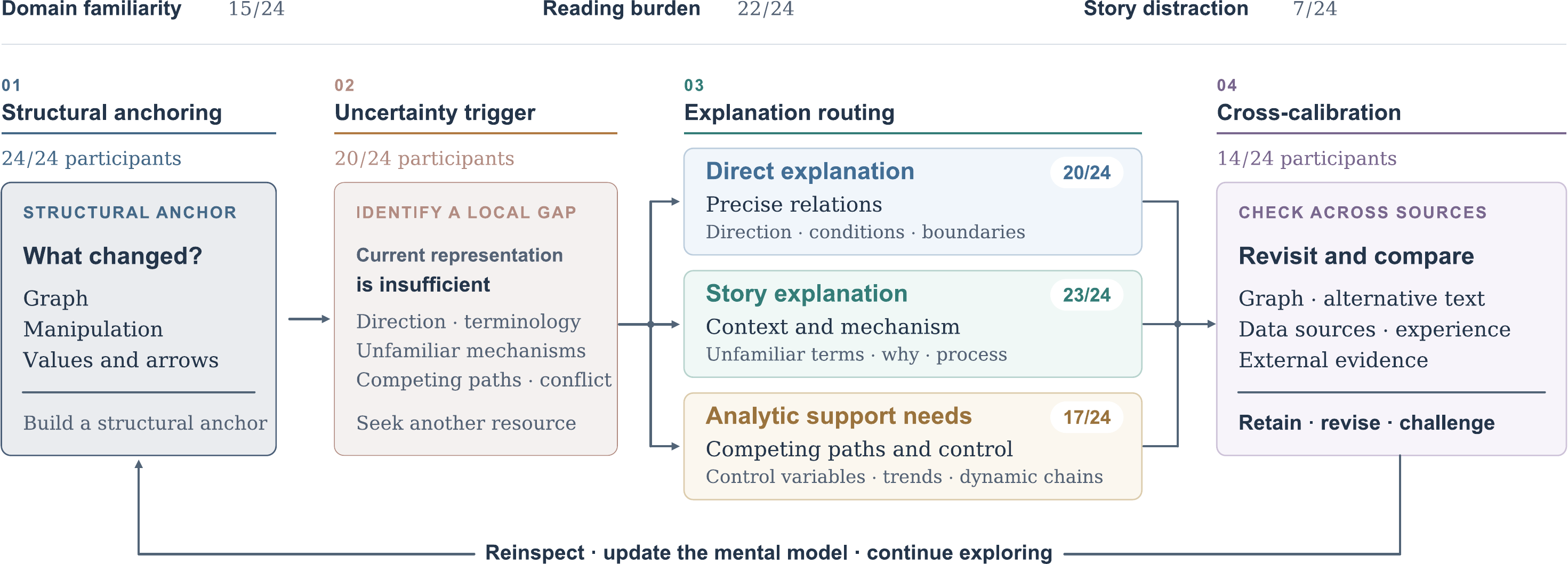}
  \caption{Adaptive causal-explanation routing framework derived from the interview coding. Participants establish a structural anchor with graphs, interaction, and values; local uncertainty triggers differentiated routing; cross-calibration yields a provisional judgment and feeds subsequent exploration.}
  \Description{A cyclical process diagram links structural anchoring, uncertainty triggering, explanation routing, and cross-calibration. Familiarity and reading cost modulate the cycle, which feeds a provisional judgment back into continued exploration.}
  \label{fig:routing-framework}
\end{figure}

We synthesize the themes as a four-stage recurring process. \emph{Structural anchoring} uses the graph, interaction, and numerical values to establish what changed. \emph{Uncertainty triggering} occurs when ambiguity about direction, terminology, mechanism, competing paths, or conflict with experience makes the current representation insufficient. During \emph{explanation routing}, users consult Direct text for precise relationships and Story for an unfamiliar context or process intuition; if competing paths remain unresolved, they seek controls, local trends, or a dynamic chain. In \emph{cross-calibration}, users return to the graph, another text, the data source, or external evidence and decide whether to retain, revise, or challenge their judgment. Familiarity and reading cost act throughout the cycle by changing when and how much explanation users reveal. This iterative retrieval, schematization, and hypothesis testing echoes classical models of analysts' sensemaking~\cite{pirolli2005}. Figure~\ref{fig:routing-framework} places information functions, triggers, moderators, and calibration outcomes within the same feedback cycle.

\begin{table}[t]
\centering
\caption{Coverage of major patterns in Study 2.}
\label{tab:themes}
\scriptsize
\begin{tabularx}{\linewidth}{>{\raggedright\arraybackslash}p{0.26\linewidth} c >{\raggedright\arraybackslash}X >{\raggedright\arraybackslash}p{0.18\linewidth}}
\toprule
Interview pattern & Coverage & Main function & Process position \\
\midrule
Structural anchoring & 24/24 & Establish a causal skeleton through graph, action, color, values, or arrows & Structural anchor \\
Direct for precise checking & 20/24 & Confirm sign, conditions, and a local chain & Precision route \\
Story for context/mechanism & 23/24 & Enter unfamiliar settings and understand why & Contextual route \\
Uncertainty-triggered switching & 20/24 & Switch after difficulty with direction, terms, mechanisms, or search & Trigger \\
Familiarity shapes choice & 15/24 & Use graph/experience in familiar domains and text in unfamiliar domains & Moderator \\
Length or terminology burden & 22/24 & Length, repetition, and jargon reduce willingness to read & Cost constraint \\
Story redundancy/distraction & 7/24 & Characters and plot add task-irrelevant information & Cost constraint \\
Competing paths/control needs & 17/24 & Competition prompts local comparisons or variable control & Trigger \\
Cross-view/external calibration & 14/24 & Check graph, text, experience, and external evidence & Cross-calibration \\
Layered/dynamic preference (probe) & 21/24 & Prefer short-first, on-demand, or dynamic presentation & Design requirement \\
\bottomrule
\end{tabularx}
\end{table}

\section{Discussion}
\label{sec:discussion}

The two studies extend the research question from a fixed comparison to the dynamic matching of explanation resources. Study 1 estimates differences between two complete texts over a predefined task set. Study 2 explains how users choose an information source in response to current uncertainty. The three resources have different costs and functions. Table~\ref{tab:integration} integrates the findings as convergence, complementarity, mechanism-oriented interpretation, or extension and separates empirical results from propositions for future testing.

\begin{table}[t]
\centering
\caption{Integration of evidence from the two studies. Interview findings explain and extend patterns from the controlled experiment.}
\label{tab:integration}
\scriptsize
\begin{tabularx}{\linewidth}{>{\raggedright\arraybackslash}p{0.24\linewidth} >{\raggedright\arraybackslash}p{0.25\linewidth} >{\raggedright\arraybackslash}p{0.13\linewidth} >{\raggedright\arraybackslash}X}
\toprule
Study 1 finding & Study 2 evidence & Relation & Integrated conclusion \\
\midrule
Positive overall accuracy difference; GEE significant & Participants chose Story for local needs while continuing to rely on graph and Direct text & Complementary & Story is useful as an on-demand resource; its average benefit is modest \\
Situational presence significantly increased & Unfamiliar terms and mechanisms often triggered Story use & Convergent & Story mainly supports entry into a context and process understanding \\
Largest benefit for total-effect adjustment & Competing paths prompted mechanism tracing, controls, and trend comparisons & Mechanism-oriented & Interviews supply a testable account of the quantitative pattern \\
Fixed texts reveal costs and benefits of each format & Free browsing reveals a cycle of anchoring, routing, and cross-checking & Extension & Interfaces should provide minimally sufficient information based on uncertainty type \\
\bottomrule
\end{tabularx}
\end{table}

\subsection{Story Supports Gaps in Context and Mechanism}

In Study 1, Story significantly increased situational presence, while differences in overall accuracy, mental simulation, and trust within the model were smaller. In Study 2, participants consulted Story around unfamiliar variables and mechanistic questions, then turned to Direct text when they needed to verify a precise relation. Together, the studies locate the principal function of Story in organizing an abstract structure into an understandable process. The graph continued to carry structural presentation, and Direct text supported precise checking. Characters and plot also increased reading cost for some participants, while domain familiarity reduced demand for Story. Story is therefore best made available on demand when a contextual or mechanistic gap emerges.

\subsection{Total-Effect Adjustment Suggests a Testable Mechanism}

The task-type interaction in Study 1 identified total-effect adjustment as the category with the clearest Story advantage. Study 2 provided a corresponding process account. When several mediating and competing paths changed at once, participants needed both to trace why variables changed and to judge which parts of the effect would be removed by fixing one variable. The sequential process in Story may help users retain the mediating chain from intervention to outcome, while Direct text helps them confirm direction, adjustment conditions, and model boundaries. This yields a directly testable mechanism hypothesis: process continuity should matter more than general immersion in total-effect tasks. A follow-up factorial experiment could manipulate process continuity and contextual immersion independently across a larger set of parallel items.

\subsection{Four Testable Propositions for Explanation Routing}

Human-centered XAI organizes explanation around users' questions~\cite{lim2009,liao2020,miller2019,wang2019,liao2021hcxai}. Our framework further specifies when an explanation appears, how much it reveals, and which representation it uses. It yields four propositions: users first establish a structural anchor; local uncertainty determines the route to an explanation; familiarity and reading cost shift the threshold for expansion; and conflict or competing paths increase the need for cross-calibration. Table~\ref{tab:propositions} maps each proposition to observable behavior and implementable interface mechanisms, making the framework suitable for system instrumentation, log analysis, and experimental testing.

\begin{table}[t]
\centering
\caption{Testable propositions from the explanation-routing framework.}
\label{tab:propositions}
\scriptsize
\begin{tabularx}{\linewidth}{>{\raggedright\arraybackslash}p{0.18\linewidth} >{\raggedright\arraybackslash}p{0.36\linewidth} >{\raggedright\arraybackslash}X}
\toprule
Proposition & Observable prediction & Interface and evaluation implication \\
\midrule
P1 Structural anchoring & Users first inspect the graph, interaction, or values and return to the structure after reading text & Persistently show the current intervention, affected nodes, and local relations; log first views and returns \\
P2 Gap-driven routing & Direction and boundary questions trigger Direct text; unfamiliar terms and mechanisms trigger Story & Recommend resources by uncertainty type; compare choices and answer revisions across triggers \\
P3 Cost modulation & With greater familiarity or longer text, users expand explanations later and read selectively & Use progressive disclosure and user control; measure expansion thresholds, reading range, and task time \\
P4 Cross-calibration & Conflict with experience and competing paths increase checks across representations, sources, or external evidence & Provide fact tracing and local comparisons; log switching sequences, evidence views, and judgment updates \\
\bottomrule
\end{tabularx}
\end{table}

\subsection{Translating the Framework into Interface Mechanisms}

\subsubsection{Keep Structure Visible and Offer a Minimal Sufficient Explanation First}

Interfaces should retain the current intervention, affected nodes, and local relationships as persistent visual anchors, with Story available as an expandable layer. A default explanation can state in one or two sentences which variable changed and through which local chain it affected the outcome. Users can then expand a longer Direct explanation or Story as needed. This form of progressive disclosure preserves the graph's navigational value while responding to the reading cost of long text. It also follows the information-visualization principle of providing an overview before details on demand~\cite{shneiderman1996}.

\subsubsection{Route Explanations by Type of Uncertainty}

Different uncertainties call for different resources. Questions about positive or negative direction, adjustment conditions, model boundaries, or task wording should route to searchable Direct text. Questions about unfamiliar terms, real-world meaning, or why a relation holds should make a contextualized Story available. When several positive and negative paths change simultaneously, the interface should support fixing other variables, highlighting a local mediator chain, comparing path contributions, or displaying a trend. Repeated viewing, switching, hovering, and follow-up questions can serve as potential uncertainty signals, while the decision to expand an explanation remains under user control.

\subsubsection{Support Cross-Checking and Contestability}

Participants checked the system against real-world experience and corroborated judgments across graphs, text, and external evidence. An explanation interface should therefore present Story as a traceable and contestable contextual scaffold. Every explanation should link back to the relevant graph relation, current value, model boundary, and data source. Fictional characters and settings added by Story should be explicitly marked as contextual devices. When users question a relationship, the system should allow them to return to the fact ledger, local structure, trend, or a more precise mechanism description.

\subsection{Evaluate Situated Experience and Task Understanding Separately}

Study 1 showed a significant increase in situational presence alongside smaller differences in overall accuracy and trust within the model. Study 2 further showed that participants who found Story useful for entering a setting could still rely on the graph and Direct text for their final judgment. Situated experience and accurate task understanding should therefore be measured as separate dimensions. Evaluations of causal explanations can jointly record objectively scored performance, subjective experience, and interaction traces such as source switching, local checking, and answer revision. Static text comparisons provide a baseline; an adaptive system must additionally test whether the timing and granularity of an explanation match the current task.

\section{Scope and Future Validation}
\label{sec:scope}

Study 1 used two reviewed instructional DAGs/SCMs and six structurally defined diagnostic questions. Its results characterize short-term judgments by everyday data users working with local causal models. The questions were generally easy, which restricted the available range for detecting group differences. With only two items per task type, task-level estimates are most useful as directional evidence. A direct next step is to expand the set of parallel total-effect-adjustment questions, increase task difficulty, and conduct a preregistered replication of the interaction.

Study 2 covered agriculture, education, and supply chains. Similar resource-use patterns recurred across 24 participants and complemented the survey results. We presented the six tasks in a fixed order to make exploratory experiences comparable. Future studies can counterbalance order and use larger samples to directly test the four process propositions.

The two texts matched on core facts, numerical values, and model boundaries, while terminology, temporal order, contextual goals, and cue salience remained part of each complete presentation package. Our estimates therefore concern the complete texts. Factorial studies can separate these components. The experience constructs were measured with single items to fit a short task session; the results characterize between-condition differences in post-task ratings. Pairing item-level confidence with accuracy would support a fuller analysis of calibration. CoNS-Explorer currently communicates supplied causal models. Future systems can retain the same interface while adding causal discovery, identification checks, and support for real-world decisions.

\section{Conclusion}
\label{sec:conclusion}

This research identifies the task boundary of Story text and the ways multiple explanation resources work together during causal-graph reasoning. In a controlled experiment with 240 participants, Story had a positive but modest association with overall accuracy, with its clearest benefit in total-effect adjustment, and significantly increased situational presence. In an independent free-browsing study with 24 participants, users established structural anchors through graphs, interaction, and numerical values; routed local uncertainty to Direct text, Story, or analytic support; and calibrated judgments against other representations and external evidence. The resulting explanation-routing framework and four testable propositions translate explanation design into responses to reasoning gaps, reading costs, and verification needs. They provide a basis for causal-explanation systems that coordinate visualization with generated text.

% Add acknowledgments only when the submission stage permits them.

\bibliographystyle{ACM-Reference-Format}
\bibliography{references}

\clearpage
\appendix
\renewcommand{\thetable}{\thesection\arabic{table}}
\renewcommand{\thefigure}{\thesection\arabic{figure}}
\section{Overview}
\label{app:overview}
\setcounter{table}{0}
\setcounter{figure}{0}
This appendix provides methodological and implementation details for the two studies in the order used by the main article. It addresses four verification questions: whether the survey data were processed consistently, whether the Direct and Story conditions were fact-matched, how CoNS-Explorer generated and checked materials, and how Study 2 connected interaction traces with interview evidence. It also separates evidence contained here from materials retained as separate anonymized companion artifacts. Participant-facing interface screenshots remain in Chinese because the study was administered in Chinese. Information that was not recorded is marked as unavailable rather than inferred.

\begin{table}[h]
\caption{What the integrated appendix contains.}
\label{tab:app-map}
\small
\begin{tabularx}{\linewidth}{p{0.22\linewidth}Y}
\toprule
Section & Evidence provided \\
\midrule
Study 1 data & Quality audit, scoring, measurement decisions, primary analysis, robustness checks. \\
Study 1 materials & Six causal-discrimination items, answer keys, challenge types, text-length matching. \\
System design & Frozen causal source, static scenario enumeration, Direct/Story generation, validation checks. \\
Study 2 & Participants, free-exploration workflow, logging/interviews, coding and framework evidence. \\
Reproducibility & Core files, commands, versioning, anonymization and known boundaries. \\
\bottomrule
\end{tabularx}
\end{table}

\section{Study 1: Survey Data Processing}
\label{app:study1-processing}
\setcounter{table}{0}
\setcounter{figure}{0}
Study 1 analyzed 240 complete survey responses, with 120 participants in Direct and 120 in Story. Six objective causal-discrimination items were scored against a frozen DAG/SCM answer key. The participant-by-item data were reshaped into 1,440 item-level observations for the primary model. Six-item total score was retained only as a robustness summary, not as the primary estimand.

\begin{table}[h]
\caption{Data-quality audit and scoring decisions.}
\label{tab:app-audit}
\small
\begin{tabularx}{\linewidth}{p{0.28\linewidth}p{0.24\linewidth}Y}
\toprule
Audit item & Result & Decision \\
\midrule
Complete responses & 240 & All complete cases entered analysis. \\
Condition allocation & Direct = 120; Story = 120 & Original randomized allocation retained. \\
Missing objective items & 0 & No imputation. \\
Missing experience ratings & 0 & No imputation. \\
Duplicate respondent/user ID & 0 & No ID-based exclusion. \\
Repeated IP flags & Some audit flags & Not used as a standalone exclusion rule. \\
Direct-only attention item & Present only in Direct & Not used for asymmetric exclusion. \\
Completion time & Conditions differed in total item count & Not used for causal comparison; no post-hoc speed cutoff. \\
Final sample & 240 & Exclusions = 0. \\
\bottomrule
\end{tabularx}
\end{table}

\paragraph{Measurement decisions.}
The six objective items were not treated as a single psychometric scale: they intentionally covered different causal operations, with two items each for confounding, outcome selection, and total-effect adjustment. Therefore we did not report Cronbach's alpha, EFA, or CFA for the six-item task. For post-task experience, the first four narrative-experience indicators had insufficient internal consistency for a composite score (Direct $\alpha=.706$, Story $\alpha=.317$, pooled $\alpha=.599$). They were therefore analyzed separately with Holm correction. Model-internal trust and cognitive effort were treated as separate planned indicators.

\section{Study 1: Main Statistical Results}
\label{app:study1-results}
\setcounter{table}{0}
\setcounter{figure}{0}
Overall task accuracy was high: raw accuracy was 87.9\% in Direct and 91.5\% in Story; 76.7\% of participants answered all six objective items correctly; four of six items had pooled accuracy at or above 90\%. This ceiling pattern motivates reporting effect sizes and uncertainty intervals rather than interpreting a single significance test in isolation.

\begin{table}[h]
\caption{Confirmatory and robustness results for overall causal-discrimination performance.}
\label{tab:app-primary}
\small
\begin{tabularx}{\linewidth}{p{0.29\linewidth}p{0.22\linewidth}p{0.24\linewidth}Y}
\toprule
Analysis & Estimate & Interval/test & Reporting role \\
\midrule
Confirmatory GLMM & OR = 1.55 & 95\% CI [0.34, 7.10], $p=.572$ & Primary participant-specific inference; no reliable overall Story advantage. \\
Marginal accuracy & Direct = 86.1\%; Story = 87.2\% & Difference = +1.16 pp, cluster-bootstrap 95\% CI [-0.48, 3.25] & Effect-size summary. \\
Participant total score & Direct = 5.28; Story = 5.49 & Difference = +0.217/6, $p=.211$, Hedges' $g=.161$ & Participant-level robustness. \\
GEE robustness & OR = 1.89 & 95\% CI [1.02, 3.48], $p=.042$ & Population-average sensitivity; not a replacement for the GLMM. \\
\bottomrule
\end{tabularx}
\end{table}

Exploratory analyses tested whether the Story effect differed across causal challenge types. The Story $\times$ Task Type interaction was supported in both frameworks: GLMM likelihood-ratio $\chi^2(2)=8.06$, $p=.018$; GEE robust Wald $\chi^2(2)=7.02$, $p=.030$. The most stable localization was that the Story effect was stronger for total-effect adjustment than for confounding (GLMM ratio OR = 7.18, Holm $p=.037$; GEE ratio OR = 2.21, Holm $p=.049$). The within-total-effect Story advantage was positive but model-sensitive (GEE: +9.08 pp, 95\% CI [3.06, 15.10], Holm $p=.009$; GLMM: OR = 6.82, Holm $p=.166$).

\begin{table}[h]
\caption{Post-task experience results. OR $>1$ indicates higher ratings under Story.}
\label{tab:app-experience}
\small
\begin{tabularx}{\linewidth}{p{0.22\linewidth}p{0.12\linewidth}p{0.12\linewidth}p{0.14\linewidth}Y}
\toprule
Outcome & Direct $M$ & Story $M$ & Ordinal OR & Result \\
\midrule
Vividness & 5.38 & 5.57 & 1.57 & Holm $p=.204$; no reliable difference. \\
Mental simulation & 5.48 & 5.64 & 1.39 & Holm $p=.339$; no reliable difference. \\
Situational presence & 5.22 & 5.77 & 2.95 & Holm $p<.001$; HC3 mean difference = +0.55, 95\% CI [0.32, 0.78]. \\
Narrative transportation & 5.88 & 5.99 & 1.20 & Holm $p=.446$; no reliable difference. \\
Model-internal trust & 5.73 & 5.57 & 0.66 & $p=.092$; not interpreted as item-level confidence calibration. \\
Cognitive effort & 4.94 & 4.88 & 0.93 & $p=.747$; means were close. \\
\bottomrule
\end{tabularx}
\end{table}

\section{Study 1: Experimental Materials}
\label{app:materials}
\setcounter{table}{0}
\setcounter{figure}{0}
The two conditions shared the same causal structures, numerical values, fixed interventions, questions, and answer keys. Direct and Story differed only in explanatory expression. Direct used compact variable-centered wording for verification; Story embedded the same causal content in a fictionalized but boundary-marked scenario. All paired texts were length-matched within 10\%: education pre-intervention 403 vs. 423 characters, education post-intervention 304 vs. 328, macro pre-intervention 441 vs. 465, and macro post-intervention 309 vs. 339.

\begin{table}[h]
\caption{Six causal-discrimination items used in Study 1.}
\label{tab:app-items}
\small
\begin{tabularx}{\linewidth}{p{0.08\linewidth}p{0.15\linewidth}p{0.58\linewidth}p{0.09\linewidth}}
\toprule
Item & Scenario & Main reasoning target & Key \\
\midrule
Q1 & Education & Negative overall association becomes positive after stratification; pre-test score is a common cause. & A \\
Q2 & Education & Conditioning on highest post-test score creates dependence through a common outcome. & D \\
Q3 & Education & Controlling self-study removes part of tutoring's total effect. & C \\
Q4 & Macro & High interest rates co-occur with high inflation because policy responds to background conditions. & B \\
Q5 & Macro & Selecting high-policy-rate years induces association through a common outcome. & B \\
Q6 & Macro & Estimating the total effect of rate increase should adjust background factors while retaining mediators. & A \\
\bottomrule
\end{tabularx}
\end{table}

\section{CoNS-Explorer System and Validation}
\label{app:system}
\setcounter{table}{0}
\setcounter{figure}{0}
CoNS-Explorer separates causal computation from language generation. A frozen causal source defines $M=\{G,S_0,a,S_1,\Delta,B,Q\}$: the DAG/SCM, baseline state, intervention, post-intervention state, changes, boundary conditions, and question set. Offline scripts enumerate scenarios, compute values deterministically, and write a shared fact ledger. The language model can read this ledger but cannot change numerical values or causal relations.

For a dataset with $k$ controllable factors, each with five forced values plus one natural state, the system enumerates $6^k$ scenarios. The public v4 interface keeps three active datasets (rice nitrogen management, tutoring education, and supply chain), producing 288 participant-facing static scenarios. Runtime interaction uses static lookup: no database, server-side LLM call, or runtime text generation is required.

\begin{figure}[h]
\centering
\includegraphics[width=.32\linewidth]{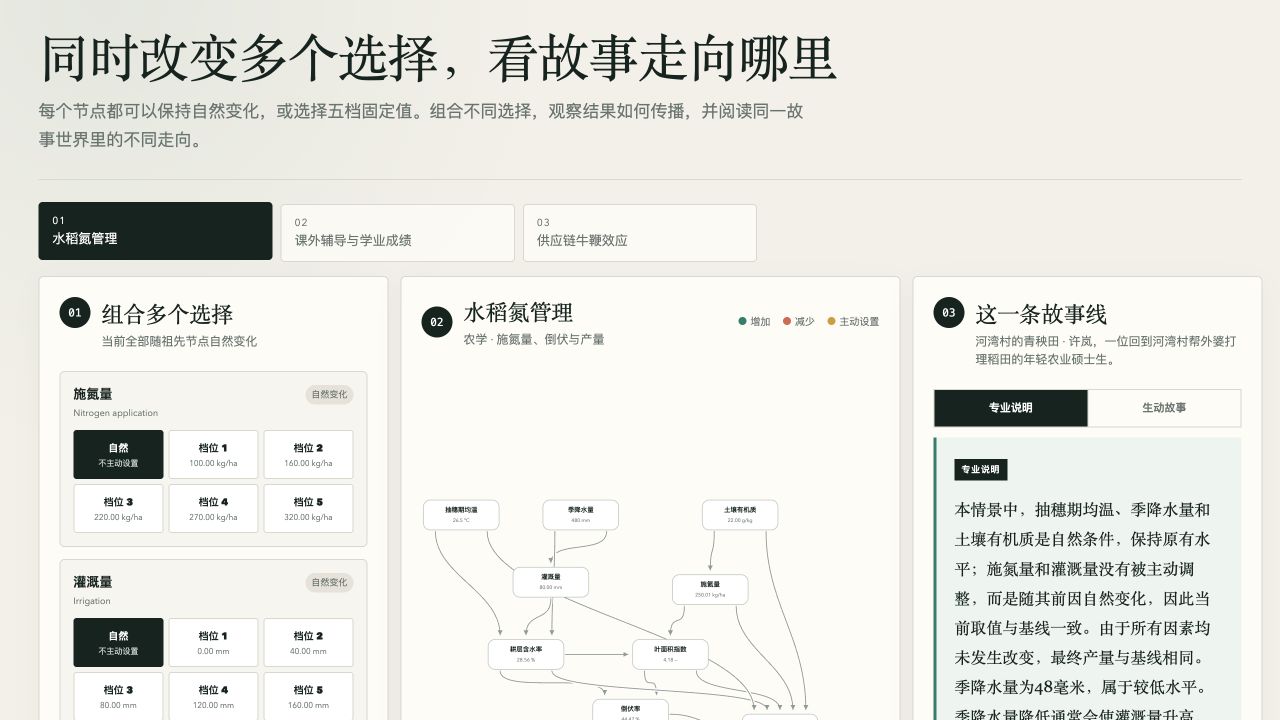}\hfill
\includegraphics[width=.32\linewidth]{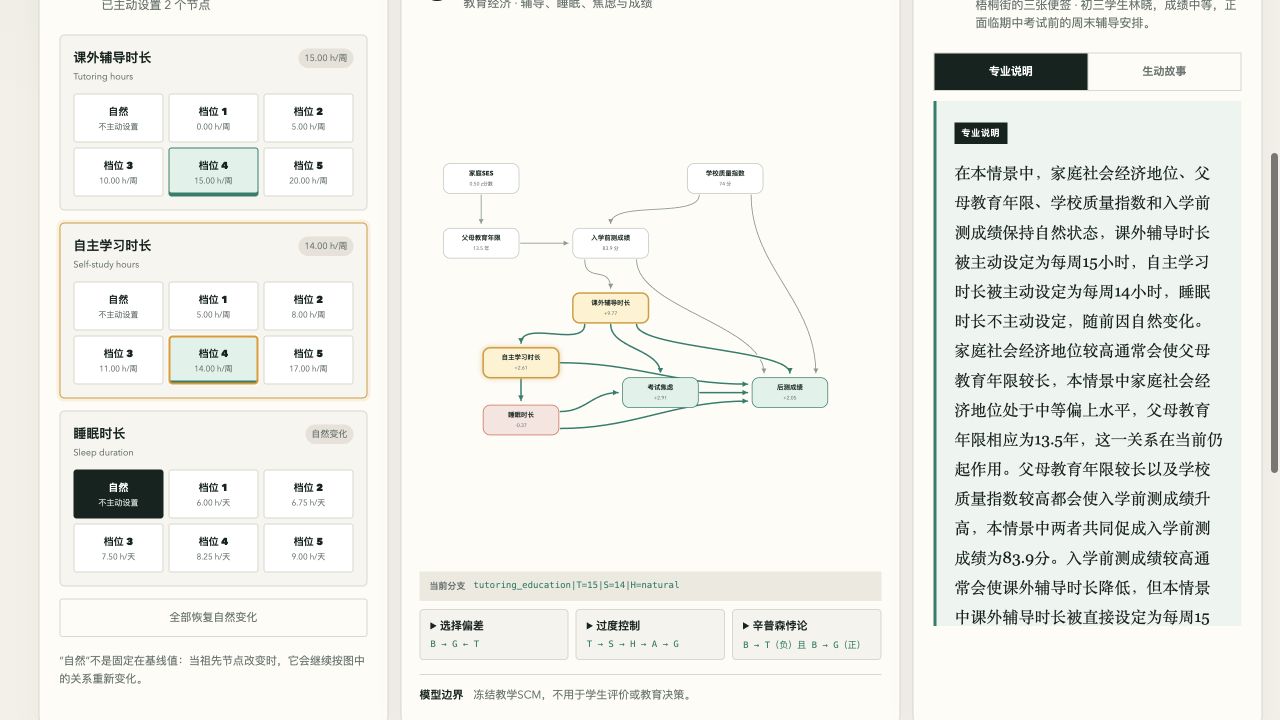}\hfill
\includegraphics[width=.32\linewidth]{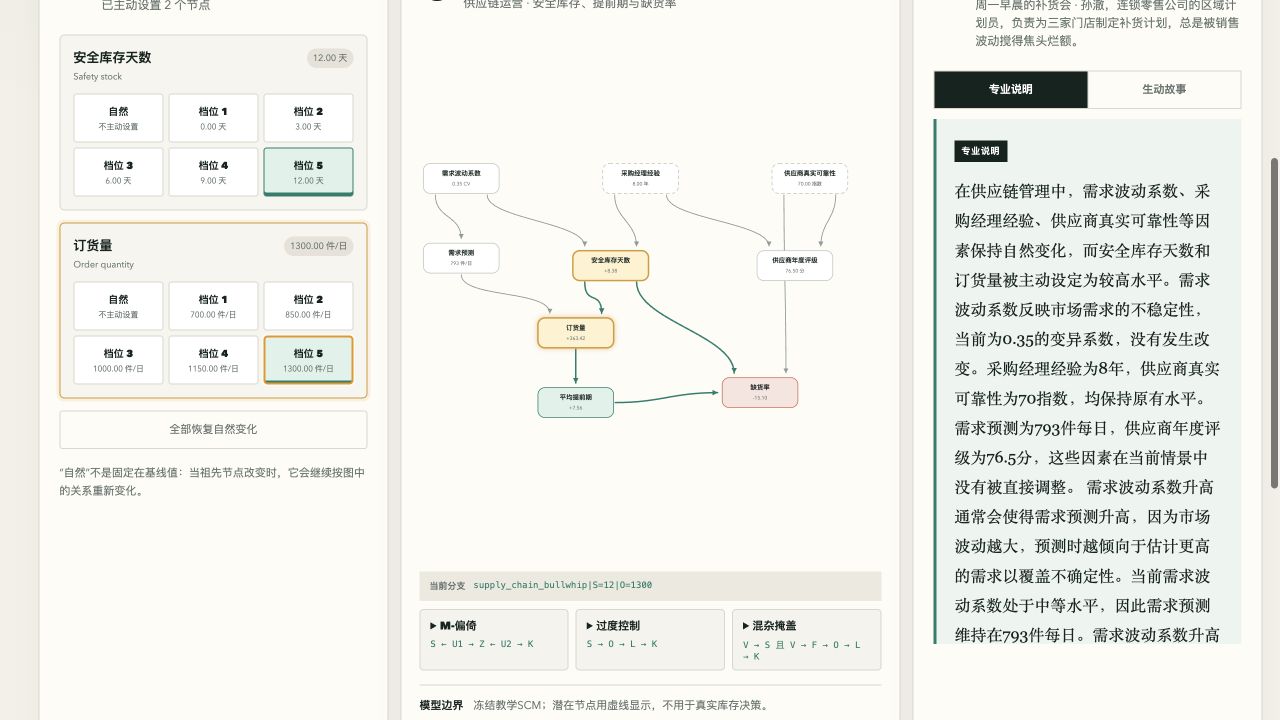}
\caption{Participant-facing Chinese interface screenshots used to document the system states: rice baseline, education with two active settings, and supply chain with two active settings. The appendix preserves the screenshots in their original language to reflect the actual study interface.}
\Description{Three screenshots of the CoNS-Explorer interface. Each shows a left control panel, a central causal graph with node changes, and a right explanation panel in Chinese.}
\label{fig:app-screenshots}
\end{figure}

\begin{table}[h]
\caption{Generation and validation controls.}
\label{tab:app-system-controls}
\small
\begin{tabularx}{\linewidth}{p{0.24\linewidth}Y}
\toprule
Component & Control \\
\midrule
Scenario enumeration & Active interventions are clamped; downstream descendants are recalculated in topological order; changed nodes use display-precision thresholds. \\
Direct explanation & Covers visible direct relations, settings, value changes, and model boundaries without internal node codes, equations, or arrows. \\
Story explanation & Uses minimal fictional context to express the same causal chain; decoration not serving causal understanding is removed. \\
Automatic checks & Unique keys, minimum text length, complete edge-coverage arrays, forbidden internal expressions, and v3/v4 story consistency. \\
Human review & Checks naturalness, fact fidelity, boundary wording, and absence of unsupported causal claims. \\
\bottomrule
\end{tabularx}
\end{table}

\section{Study 2: Exploratory Workflow and Qualitative Analysis}
\label{app:study2}
\setcounter{table}{0}
\setcounter{figure}{0}
Study 2 used an independent sample of 24 participants who had not joined Study 1. Participants were recruited through social-media channels, were at least 20 years old, had undergraduate-level education or above, and reported no more than introductory experience with causality/statistics/data analysis. Each session used a PC interface and lasted about 40 minutes including system tasks and interview. Participants completed six fixed-order tasks across three domains while freely switching among graph, Direct, and Story views.

\begin{table}[h]
\caption{Study 2 data sources and qualitative workflow.}
\label{tab:app-qual}
\small
\begin{tabularx}{\linewidth}{p{0.24\linewidth}Y}
\toprule
Element & Description \\
\midrule
Interaction data & Graph visits, Direct/Story view switches, control settings, and task completion traces where available. \\
Verbal data & Think-aloud comments and semi-structured interviews; recordings were AI-transcribed and then manually checked against audio/video. \\
Coding procedure & LLM-assisted first-pass organization was used only as a support step. Two researchers reviewed, merged, split, and renamed codes before final interpretation. \\
Core themes & Structure anchoring; uncertainty-triggered switching; complementary affordances of graph/direct/story; cross-calibration across representations or external knowledge. \\
Evidence standard & Coverage counts describe the interview sample only; they are not population proportions or hypothesis tests. \\
\bottomrule
\end{tabularx}
\end{table}

The resulting process framework is: structure anchoring $\rightarrow$ uncertainty trigger $\rightarrow$ explanation routing $\rightarrow$ cross-calibration. Graphs provide the default structural anchor. Direct explanation supports local verification of edges, conditions, and boundary wording. Story explanation supports contextual mechanism understanding when domains or processes feel unfamiliar. Cross-calibration occurs when participants return to the graph, compare text forms, or invoke external knowledge before deciding.

\section{Reproducibility, Anonymization, and Boundaries}
\label{app:reproducibility}
\setcounter{table}{0}
\setcounter{figure}{0}
The planned anonymized reproducibility package includes the frozen causal models and intervention plans; scripts for scenario generation, validation, and analysis; static scenario tables and narrative assets; questionnaire materials and anonymized survey outputs; and the final codebook. Typical commands are \texttt{npm install}, \texttt{npm run build}, \texttt{npm run scenarios:generate}, and \texttt{npm run narratives:v4:validate}. Separate R scripts reproduce the GLMM, GEE, bootstrap summaries, and ordinal and HC3 experience analyses.

All companion files will be checked for anonymity before CHI review. File names, document metadata, screenshots, comments, repository links, recruitment records, and ethics descriptions should not reveal author or institution identity. Known boundaries are also explicit: the causal models are teaching models rather than estimates from real observational datasets; discrete intervention values do not exhaust continuous spaces; Story is evaluated as a full expression package; post-task trust is not item-level confidence calibration; and qualitative counts from Study 2 describe this sample rather than a population rate.

\end{document}